\documentclass[conference]{IEEEtran}

\usepackage{float}
\usepackage{amsmath,amssymb,amsfonts}
\usepackage{mathtools}
\usepackage{algorithmic}
\usepackage[style=ieee]{biblatex}
\usepackage{graphicx}
\usepackage{textcomp}
\usepackage{optidef}
\usepackage{xcolor}
\usepackage{caption}
\usepackage{subcaption}
\DeclarePairedDelimiter\abs{\lvert}{\rvert}
\DeclareMathOperator{\Tr}{Tr}
\newcommand{\mbf}[1]{$\mathbf{#1}$}

\def\BibTeX{{\rm B\kern-.05em{\sc i\kern-.025em b}\kern-.08em
    T\kern-.1667em\lower.7ex\hbox{E}\kern-.125emX}}

\newcommand\blfootnote[1]{
  \begingroup
  \renewcommand\thefootnote{}\footnote{#1}
  \addtocounter{footnote}{-1}
  \endgroup
}
\begin{document}

\title{Generative Deep Synthesis of MIMO Sensing Waveforms with Desired Transmit Beampattern}

\author{\IEEEauthorblockN{Vesa Saarinen, Robin Rajam\"{a}ki, and Visa Koivunen}
\IEEEauthorblockA{\textit{Department of Information and Communications Engineering}, Aalto University, Finland}
}

\maketitle
\blfootnote{This work is supported in part by the following projects: Business Finland 6G-ISAC, Research Council of Finland FUN-ISAC (359094), EU Horizon INSTINCT (101139161) and SAAB Cognitive Radar.}
\begin{abstract}
This paper develops a generative deep learning model for the synthesis of multiple-input multiple-output (MIMO) active sensing waveforms with desired properties, including constant modulus and a user-defined beampattern. The proposed approach is capable synthesizing unique phase codes of on-the-fly, which has the potential to reduce interference between co-existing active sensing systems and facilitate Low Probability of Intercept/Low Probability of Detection (LPI/LPD) radar operation. The paper extends our earlier work on synthesis of approximately orthogonal MIMO phase codes by introducing flexible control over the transmit beampatterns. The developed machine learning method employs a conditional Wasserstein Generative Adversarial Network (GAN) structure. The main benefits of the method are its ability to discover new waveforms on-demand (post training) and generate demanding beampatterns at lower computational complexity compared to structured optimization approaches.
\end{abstract}

\section{Introduction}

Modern radars and active sensing systems deploy digital, fully adaptive large aperture sensor arrays both at the transmitter and receiver. These systems may perform multiple functions requiring different waveforms, which may be launched from different radiating elements. Large aperture arrays used in cognitive multiple-input multiple-output (MIMO) radars, massive MIMO and millimeter wave communications systems may use a very large number of antennas and independent waveforms simultaneously. Various sensing tasks require different beampatterns in the spatial domain, and the transmitter should be able to form complicated beampatterns that may not be feasible using conventional phased arrays. 

Commonly-used radar code families such as Barker codes have relatively few codes which reduces flexibility, the applicability in large arrays, and can be detrimental in adversarial operational environments. As such, novel waveform synthesis approaches 
are of great interest to help explore the high-dimensional space of complex waveforms and discover new waveform families with desirable properties. To this end, we apply a generative deep learning approach to the synthesis and discovery of constant-modulus waveforms that produce desired beampatterns with high fidelity.

Structured optimization approaches to the problem exist as well \cite{Aittomaki_Synthesis,Stoica_Synthesis}. These approaches can consider a variety of factors, such as peak sidelobe level, cross- and auto-correlation properties, unit modulus or low peak-to-average-power-ratio constraints. While optimization approaches produce high-quality waveforms, their computational cost can be prohibitive for on-demand waveform generation, and the computational time needed may vary greatly depending on the constraints and initialization. This is especially the case as the desired number of codes and their length increases. 

This paper extends our previous generative deep learning work on orthogonal MIMO radar waveform generation \cite{Paper3} through the synthesis of MIMO phase codes with arbitrary cross-correlation matrices. This allows for controlling transmit beampatterns in MIMO sensing applications, and producing demanding beamshapes that cannot be obtained using conventional phased arrays. 
While the training of deep learning models can be time-consuming, producing outputs with a trained model is generally fast with a predictable computation time. These properties are appealing for on-demand waveform synthesis. The proposed method employs a Generative Adversarial Network (GAN) \cite{GAN} model \emph{conditioned} on the cross-correlation matrix of the transmit waveforms. After training, the approach can synthesize novel waveforms considerably faster than conventional optimization approaches.

The paper is organized as follows. Section \ref{sec:GAN} briefly introduces the relevant background on GANs. Section \ref{sec:approach} then presents the proposed generative model, including its structure, penalty terms, and details of training data generation. Section \ref{sec:results} numerically illustrates the advantages of the generative model. Finally, section \ref{sec:conclusion} concludes the paper.

\section{Basics of Generative Adversarial Networks}\label{sec:GAN}

\begin{figure}
    \centering
    \includegraphics[width=0.7\linewidth]{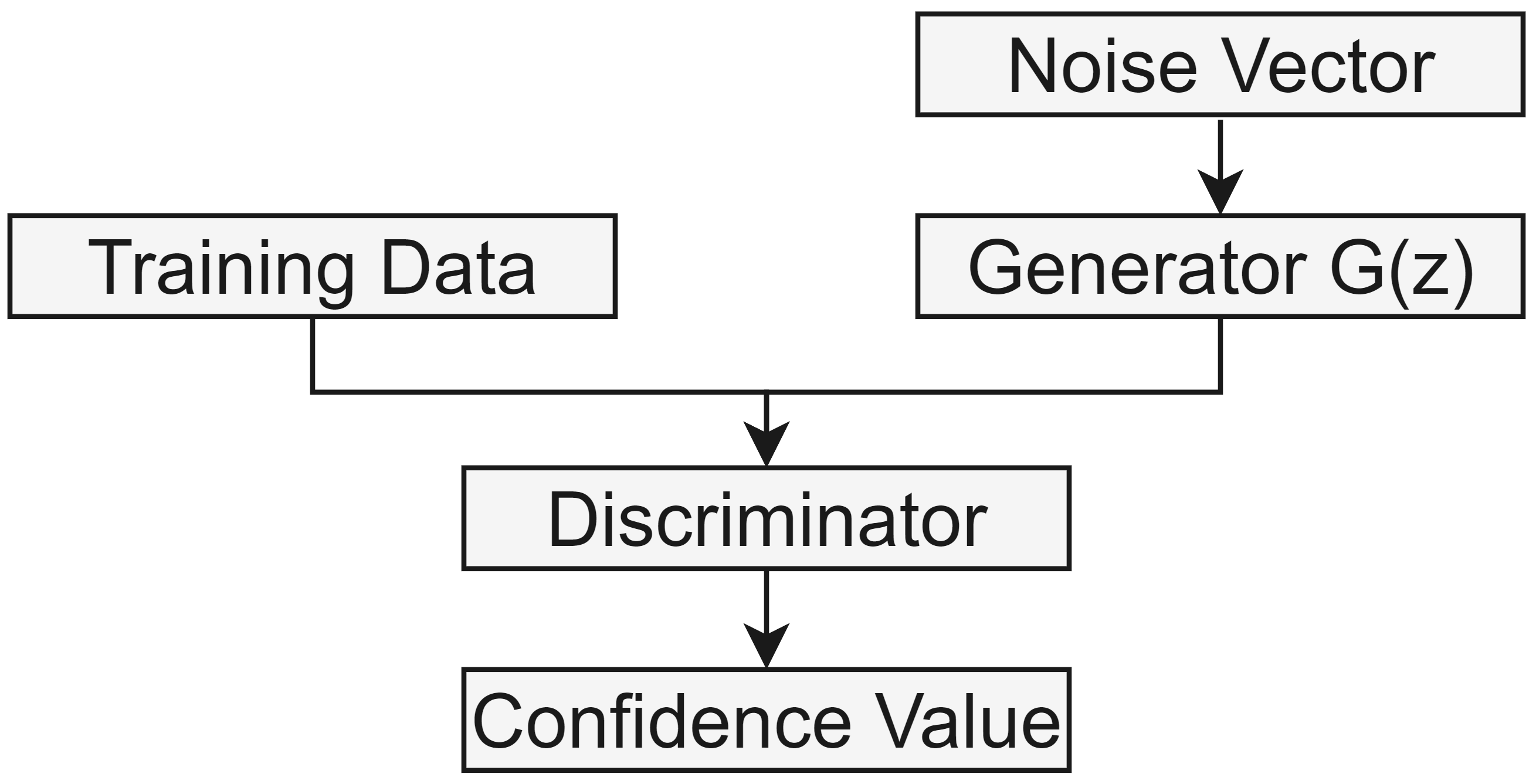}
    \caption{The basic GAN structure. The generator receives a noise vector as input, and produces fake data samples. The discriminator assigns a confidence value to its inputs, attempting to separate real training data from fake samples.}
    \label{fig:GAN_graph}
\end{figure}

Generative Adversarial Networks (GAN) \cite{GAN} are deep learning models that implicitly learn to draw samples from a data distribution. A GAN is comprised of two networks, a generator $G(\mathbf{z})$ and a discriminator $D(\mathbf{x})$. This structure is illustrated in Figure \ref{fig:GAN_graph}. The task of the discriminator is to classify whether its inputs, waveforms in our case, are drawn from a training data set, or if they have been produced by the generator. The generator receives as its input a noise vector, and aims to produce samples that fool the discriminator. As both networks improve their performance through training, the generator implicitly learns the distribution of the training data (without directly ``seeing'' it). The generator can then be used to draw samples from the learned distribution. In our design problem it means generating a matrix of phase codes. 

A variety of GAN formulations exist \cite{WGAN,LSGAN,CGAN}. In this paper, we use Wasserstein GANs (WGAN) \cite{WGAN}, which minimize the earth-mover (EM) distance between the generated and training data distributions. Essentially, this is the probability mass times distance that would need to be moved in order to transform one of the distributions into the other. This is a useful notion of distance, since it has a gradient even for disjoint distributions. As such, WGAN models feature an unbounded discriminator confidence value $D(\mathbf{x})$ to produce a useful gradient for generator training even in these scenarios. The primary advantage of this formulation is high training stability. The primary drawback is a relatively high computational cost during training. This is caused by the need for additional normalization as well as over-training the discriminator \cite{WGAN}. The latter refers to training the discriminator for multiple batches for each batch of generator training. Typically, the WGAN loss term is defined as
\begin{equation}
\begin{split}
\min_{G}\max_{D}\;&\mathbb{E}_{\mathbf{x}\sim p_{d}}[D(\mathbf{x})]-\mathbb{E}_{\mathbf{z}\sim{p_{z}}}[D(G(\mathbf{z}))],
\end{split}
\end{equation}
where $p_{d}$ is the distribution of data samples, $p_{z}$ is the generator noise distribution, which is most often standard normal.
WGAN training additionally requires Lipschitz normalization to keep the gradient of the discriminator close to 1 \cite{WGAN}. This is done to make the discriminator loss a good representation of EM distance. In the proposed method we apply Gradient Penalty (GP) \cite{WGAN2} and the WGAN loss function is modified to
\begin{equation}
\begin{split}
\label{WGAN}
\min_{G}\max_{D}\;& \mathbb{E}_{\mathbf{x}\sim p_{d}}[D(\mathbf{x})]-\mathbb{E}_{\mathbf{z}\sim p_{z}}[D(G(\mathbf{z}))]\\
&\quad+ \lambda \mathbb{E}_{\mathbf{s}\sim p_{s}}[(\abs{\nabla_{s}D(\mathbf{s})}_{2}-1)^2],
\end{split}
\end{equation}
where $p_{s}$, which penalty is applied on, is a random distribution of points in the sample space. Ideally, gradient penalty would be applied across the entire space. However, this is not possible in a continuous space. As such, it is most often evaluated on points that lie between the training data distribution and generator output distribution. During training, the generator distribution will ideally move towards the training data distribution, and as such points between the two are of interest.

GANs can also use conditioning variables \cite{CGAN}, such as class labels in their inputs. This can apply to either of the neural networks or both. Class labels allow for the learning of conditional distributions, such as in our case producing a desired cross-correlation matrix based on a conditioning input.

\section{Proposed generative MIMO phase code synthesis with desired transmit beampattern}
\label{sec:approach}

\begin{figure}
    \centering
    \includegraphics[width=1\linewidth]{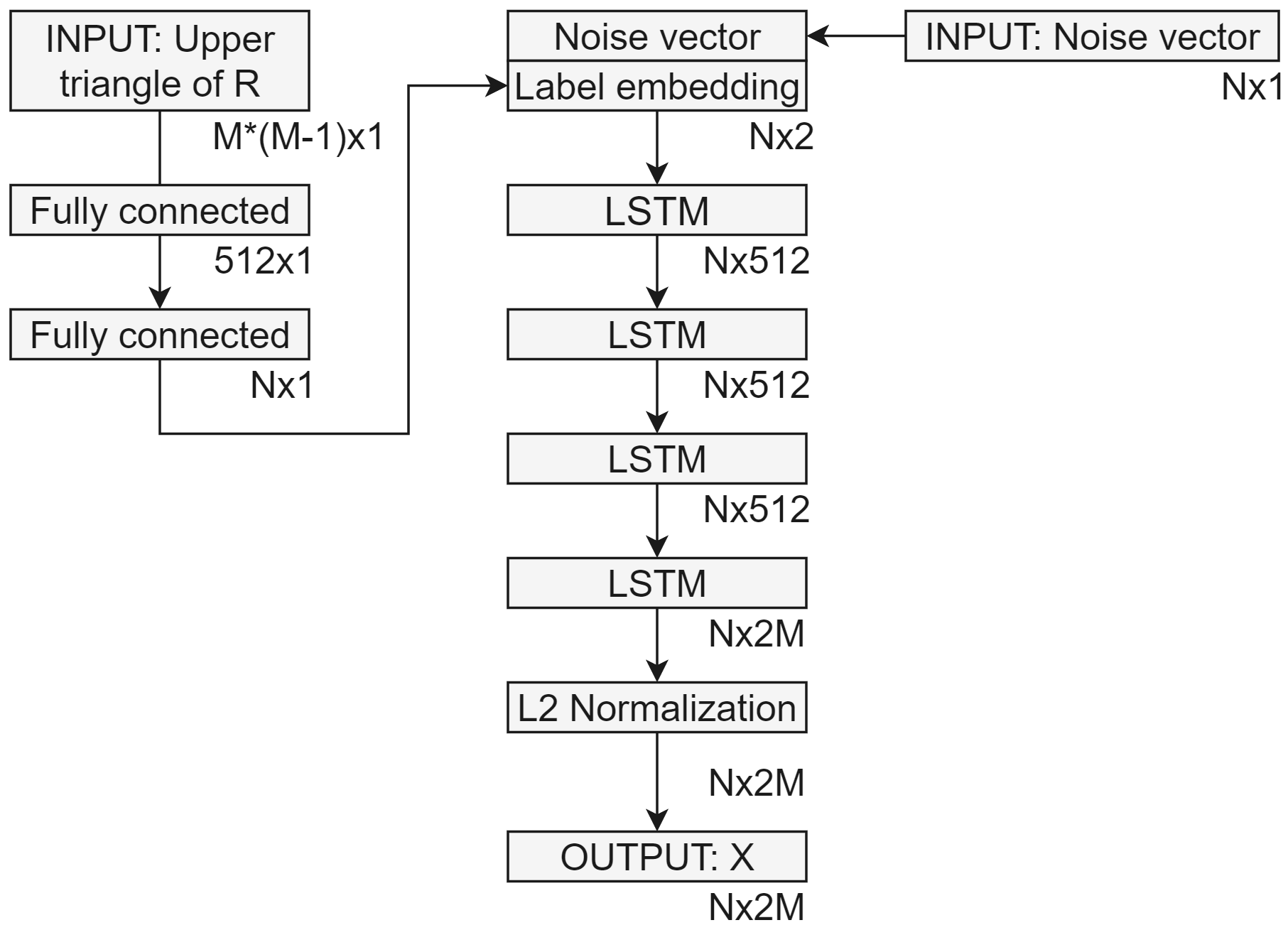}
    \caption{Proposed conditional generator network structure, with the output dimensions of each layer indicated. The generator is conditioned using cross-correlation matrix \mbf{R} to produce code matrices \mbf{X} satisfying $\mathbf{X}^H\mathbf{X}=\mathbf{R}$ (approximately).}
    \label{fig:Generator_Structure}
\end{figure}

We propose a GAN-based constant-modulus waveform synthesis approach for generating phase code matrices $\mathbf{X}\in\mathbb{C}^{N\times M}$ with arbitrary positive semi-definite cross-correlation matrices  $\mathbf{R}\in\mathbb{C}^{M\times M}$. The generator structure is illustrated in Figure \ref{fig:Generator_Structure}. This section introduces the generative model and related penalties.
Our goal is to learn a mapping $G(\mathbf{z, r}) : \mathbb{R}^N\times \mathbb{R}^{M(M-1)} \mapsto \mathbb{C}^{N\times M}$ such that each input label vector $\mathbf{r}\in\mathbb{R}^{M(M-1)}$ corresponds to a distribution of waveform matrices \mbf{X}. All the matrices in a distribution should (approximately) satisfy $\mathbf{X}^H\mathbf{X} =\mathbf{R}$. The vector $\mathbf{r}$ is constructed from the upper triangular part of \mbf{R}, and hence uniquely defines the matrix. By learning to produce a specific \mbf{R}, we also learn to produce transmit beampattern $b(\theta)=\mathbf{a}(\theta)^H\mathbf{R}\mathbf{a}(\theta)$, where $\mathbf{a}(\theta)\in\mathbb{C}^{M}$ is the steering vector corresponding to angle $\theta$ of the given array geometry. Note that the generative model actually has the benefit of being agnostic to the choice of array geometry, since it is conditioned on \mbf{R} rather than $b(\theta)$. The other input, $\mathbf{z}\in\mathbb{R}^{N}$, is a standard normal noise vector. 

\subsection{Generator}
The generator, illustrated in Figure \ref{fig:Generator_Structure}, is a conditional LSTM \cite{LSTM} network. LSTM networks are recurrent networks that process their inputs and outputs in a sequential manner. These networks capture relationships between subsequent elements while retaining some information over longer periods, and are most often applied to time-series data. They are our choice of architecture due to the sequential structure of phase codes. The generator architecture was empirically found to strike a balance between computational cost of training and post-training performance. An LSTM produces an output element for each input element. To produce a sequence of $N$ outputs, where each element of the sequence is an $1\times M$ row of the code matrix, we shape our inputs to be of length $N$

For the conditioning input $\mathbf{r}$, we use the upper triangle of the $M\times M$ cross-correlation matrix $\mathbf{R}$, excluding the diagonal. As $\mathbf{R}$ is Hermitian, and every diagonal entry equals $1$ ($\mathbf{X}$ is constant-modulus), this upper triangle uniquely defines $\mathbf{R}$. The upper triangular entries are unfolded into an $M(M-1)/2\times1$ vector, and split into real and imaginary components. The two real-valued vectors are then concatenated to produce an $M(M-1)\times1$ input label vector \mbf{r}. A learned $N\times1$ embedding of the input conditioning vector $\mathbf{r}$ using two fully connected layers is combined with the $N\times1$ generator noise vector to form an $N\times2$ input for the  generator LSTM layers. 

In order for our network to operate on real numbers instead of complex numbers, we produce $N\times2M$ waveform matrices, where each adjacent column pair is a bivariate real representation of a complex unit-modulus waveform. These matrices are then transformed into an $N\times M$ complex representation. To ensure constant modulus, the generator output is normalized such that each complex entry $\abs{G(\mathbf{z, r})_{i,j}}=1$. 

\subsection{Discriminator}
For the discriminator $D(\mathbf{z, r})$, we define a conditional convolutional structure in Table \ref{ConvDis}. The discriminator uses a conditional structure similar to the generator, with \mbf{r} having a learned embedding through a fully-connected layer before being combined with the input code matrix \mbf{X}. The discriminator generally does not need to be as powerful as the generator. As such the number of parameters for this network is lower, and we use a convolutional structure which is generally computationally faster. The convolutional network operates on data through convolution with different sizes of filters moving across the input in steps with a certain stride length. The discriminator in our model allows for the implicit learning of phase structure from the training data. This can include desirable properties such as ambiguity function defining delay and Doppler response. It also encourages diversity in outputs, as a generator which has its outputs confined to a small subset of the training data is easily penalized by the discriminator.

We additionally employ Gaussian noise with a standard deviation of 0.1 in the discriminator input to improve training stability when training with limited amounts of data. As the noise is applied to both the training data and generator data distributions, this does not affect the distribution learned by the generator \cite{GAN_Training_Limited_Data}. The amount of noise applied is determined experimentally to achieve the desired stabilization without heavily degrading discriminator performance.

\begin{table}[h!]
\centering
\caption{Discriminator structure. The discriminator is a conditional convolutional network with LeakyReLu activations. The input label vector \mbf{r} is fed through a fully-connected layer and combined with the input code matrix \mbf{X} (augmented by Gaussian noise). Convolution outputs for our data, $N=41, M=10$. The model outputs a scalar confidence value.}
\begin{tabular}{||c c c c||} 
\hline
Operation & Kernel Size & Strides & Output Shape \\ [0.5ex] 
\hline\hline
Input (\mbf{r}) & - & - & (M(M-1)) \\ 
Dense \& Reshape & - & - & (N, M, 1) \\ 
Input (\mbf{X}) & - & - & (N, M, 2) \\ 
Gaussian Noise & - & - & (N, M, 2) \\ 
Concat & - & - & (N, M, 3) \\ 
Conv2D & (5,5) & (2, 1) & (20, 8, 128) \\
Conv2D & (4,4) & 2 & (10, 4, 256) \\
Conv2D & (4,4) & 2 & (5, 2, 512) \\
Fully Connected & - & - & (1) \\  [1ex] 
\hline
\end{tabular}
\label{ConvDis}
\end{table}

\subsection{Loss function}
In order to produce the desired cross-correlation matrix $\mathbf{R}$, we define an additional penalty term for the generator. While this property can be learned implicitly, consistency in global structure is often a challenge for generative deep learning methods. As such, an explicit penalty can help ensure the desired \mbf{R} is produced, especially as the size of generated samples increases. We define the penalty term as:
\begin{equation}
l=\abs{\abs{N\mathbf{R} - G(\mathbf{z}, \mathbf{r})^HG(\mathbf{z}, \mathbf{r})}}_F.
\label{CorrelationPenalty}
\end{equation}
The penalty term uses a desired cross-correlation matrix \mbf{R} and the associated conditioning vector \mbf{r} to penalize the model for producing waveforms whose cross-correlation matrix does not closely approximate the desired \mbf{R}. Applying this penalty with a variety of choices of $\mathbf{R}$ during training allows the generator to learn this association for different conditioning inputs. The correlation penalty term is multiplied by scalar $\nu \geq 0$ and added to the usual WGAN loss term in (\ref{WGAN}):
\begin{equation}
\begin{aligned}
\min_{G}\max_{D} \;&
 \mathbb{E}_{\mathbf{x}\sim p_{d}}[D(\mathbf{x})]-\mathbb{E}_{\mathbf{z}\sim{p_{z}}}[D(G(\mathbf{z}))- \\
 &\lambda \mathbb{E}_{\mathbf{s}\sim p_{s}}[(\abs{\abs{\nabla_{s}D(\mathbf{s})}}_{2}-1)^2]+ \nu l].
\end{aligned}
\end{equation}
As the WGAN loss for the discriminator is unbounded, $\nu$ has to be determined experimentally and depends on the dataset, training process, and discriminator structure.

\subsection{Training data generation} \label{sec:training}

Generation of phase codes providing an approximation of a specific transmit beampattern $\mathbf{a}(\theta)^H\mathbf{R}\mathbf{a}(\theta)$ has been widely explored in literature. We use an optimization method related to \cite{Stoica_Synthesis} and \cite{Aittomaki_Synthesis} to produce training data for our waveform synthesis approach. The proposed approach is comprised of two steps. First, we produce a cross-correlation matrix \mbf{R} from the desired beampattern. Then, we generate a code matrix \mbf{X} based on the cross-correlation matrix. This allows us to design training code sets that produce the desired beampattern with high fidelity when launched from co-located MIMO radar antenna elements. 

For the first step, we use a Hermitian square root matrix form for the correlation matrix $\mathbf{R} = \mathbf{L}\mathbf{L}^H$ to ensure positive semi-definiteness of \mbf{R}. Then, we optimize for:
\begin{mini*}|s|
{\alpha, \mathbf{L}}{\dfrac{1}{I}\sum_{i=1}^{I}[\alpha b(\theta_i) - \mathbf{a}(\theta_i)^H\mathbf{L}\mathbf{L}^H\mathbf{a}(\theta_i)]^2}
{}{}
\addConstraint{\sum_{m=1}^{M}\abs{L_{nm}}^2} = 1{}\; \text{for} \; n=1,...,N
\end{mini*}
where $b(\theta)$ is the desired transmit beampattern. This will produce a positive semi-definite cross-correlation matrix that gives approximation of the desired beampattern in the least squares sense. With the row norm constraint on \mbf{L}, matrix $\mathbf{R}$ will have ones along the diagonal and (complex-valued) nondiagonal entries of magnitude between 0 and 1. The optimization problem is formulated using a real-valued vector representation of \mbf{L} and a local optimum is found using the SciPy optimize solver. We then use the MultiCAO algorithm as described in \cite{Stoica_Synthesis} with random initializations to create training data code matrices $\mathbf{X}$ from $\mathbf{R}$. 

In the results presented here, we produce batches of $M=10$ waveforms of length $N=41$. During training, we train the discriminator for 5 iterations for each iteration of generator training. Gradient penalty is evaluated on points that are randomly weighted averages between pairs of training data samples and generator outputs belonging to the same class in the discriminator training batch: $t\mathbf{X}+(1-t)G(\mathbf{z,r})$, $t\in[0,1]$ is a uniform random variable. We use a correlation penalty coefficient of $\nu = 10$ and gradient penalty coefficient of $\lambda=10$. 

\section{Simulation Results} \label{sec:results}

\begin{figure}
    \centering
    \begin{subfigure}[b]{0.85\linewidth}
        \centering
        \includegraphics[width=\linewidth]{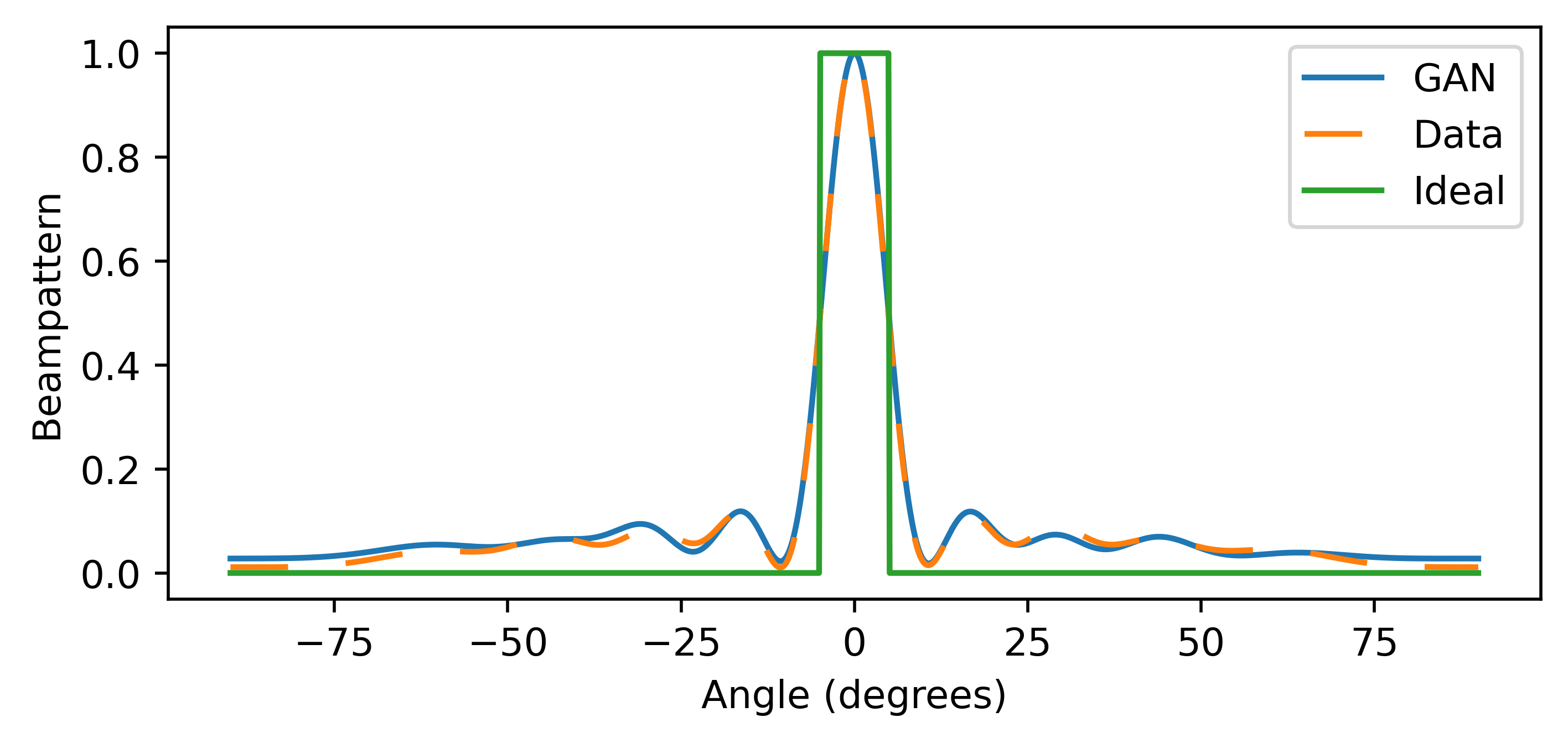}
    \end{subfigure}
    
    \begin{subfigure}[b]{0.85\linewidth}
        \centering
        \includegraphics[width=\linewidth]{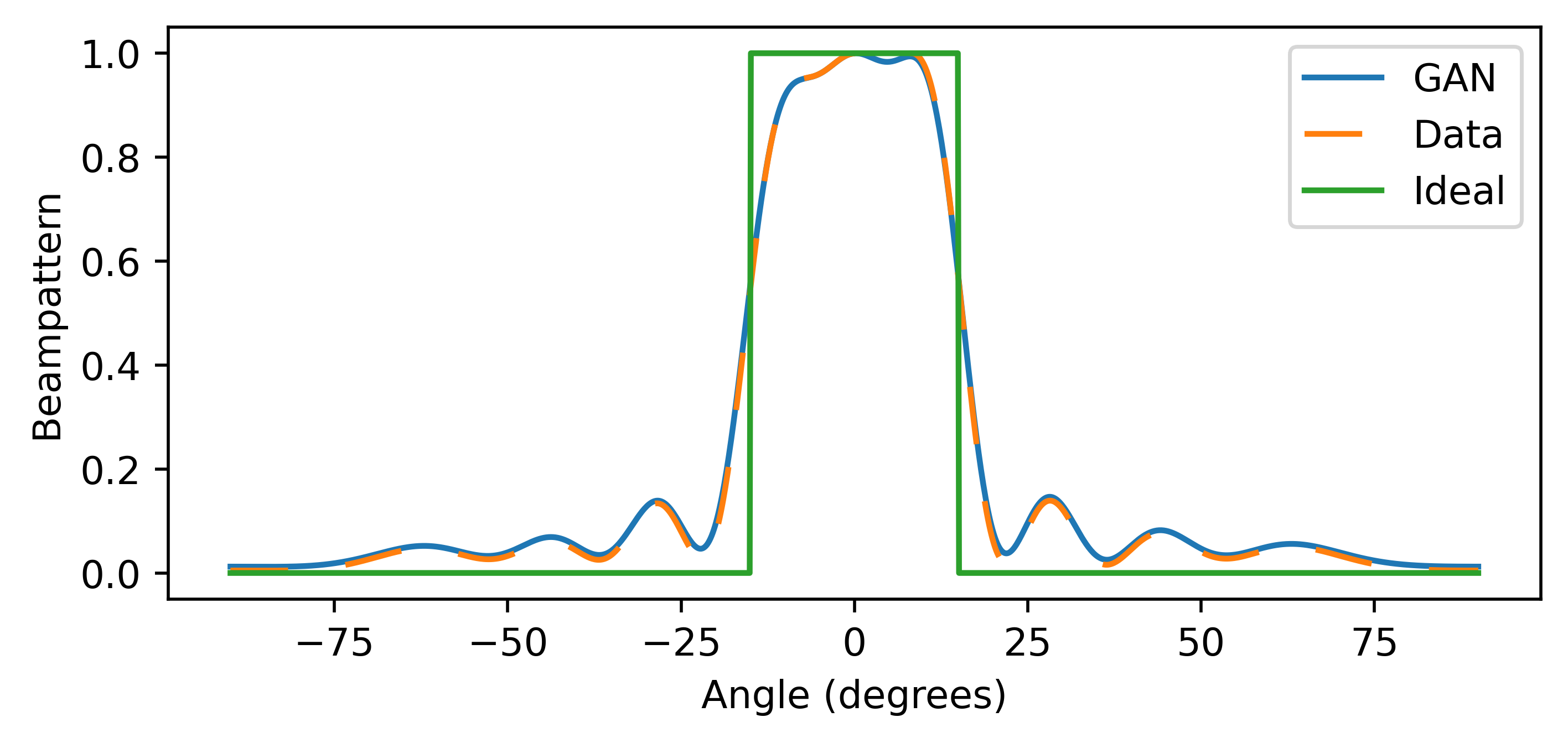}
    \end{subfigure}

    \begin{subfigure}[b]{0.85\linewidth}
         \centering
         \includegraphics[width=\linewidth]{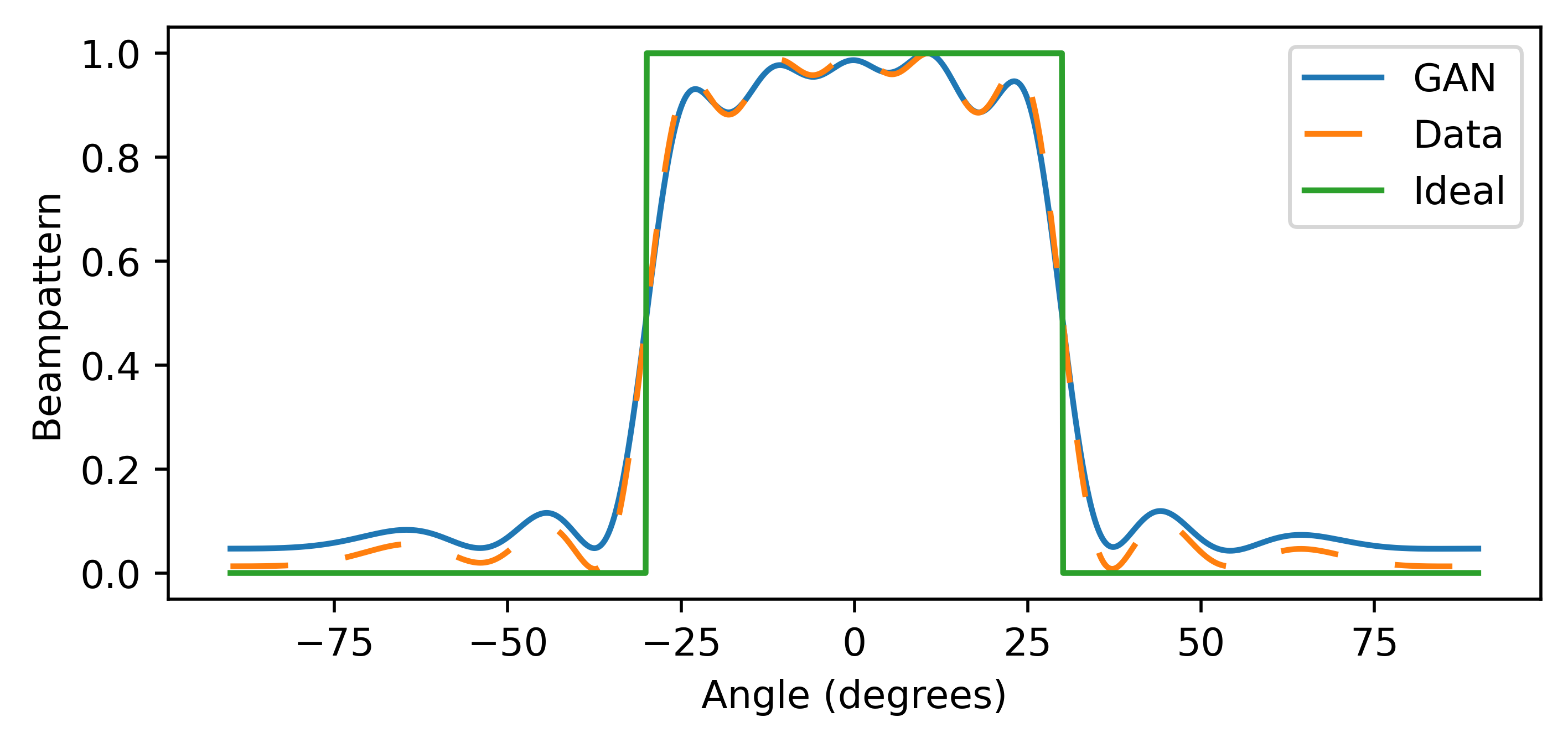}
    \end{subfigure}
    
    \begin{subfigure}[b]{0.85\linewidth}
         \centering
         \includegraphics[width=\linewidth]{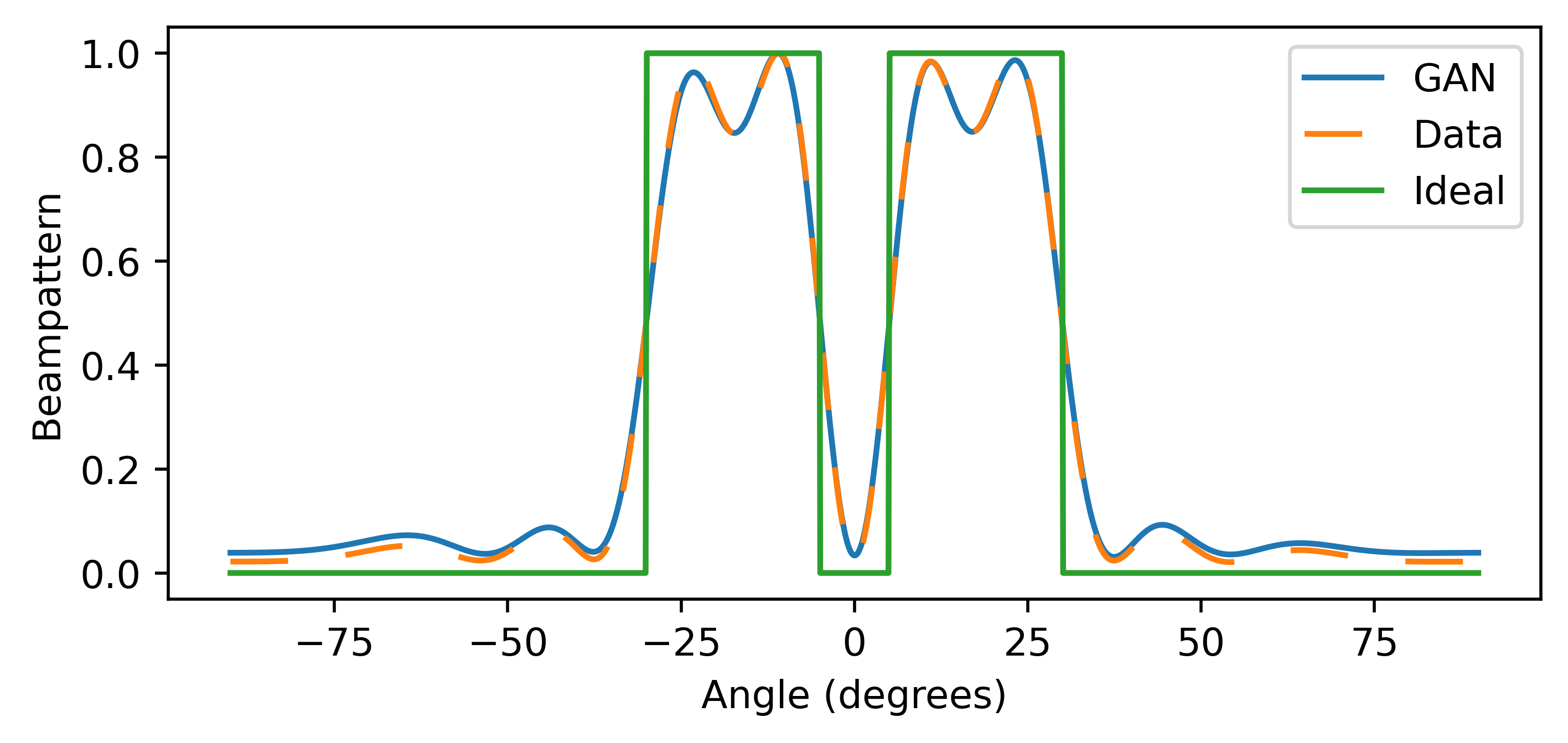}
     \end{subfigure}
        \caption{Beampatterns of generative model for different beam shapes. The GAN learns to synthesize novel waveforms that accurately mimic the training data beampatterns.}
        \label{fig:gan_bp}
\end{figure}

\begin{figure}
    \centering
    \includegraphics[width=0.8\linewidth]{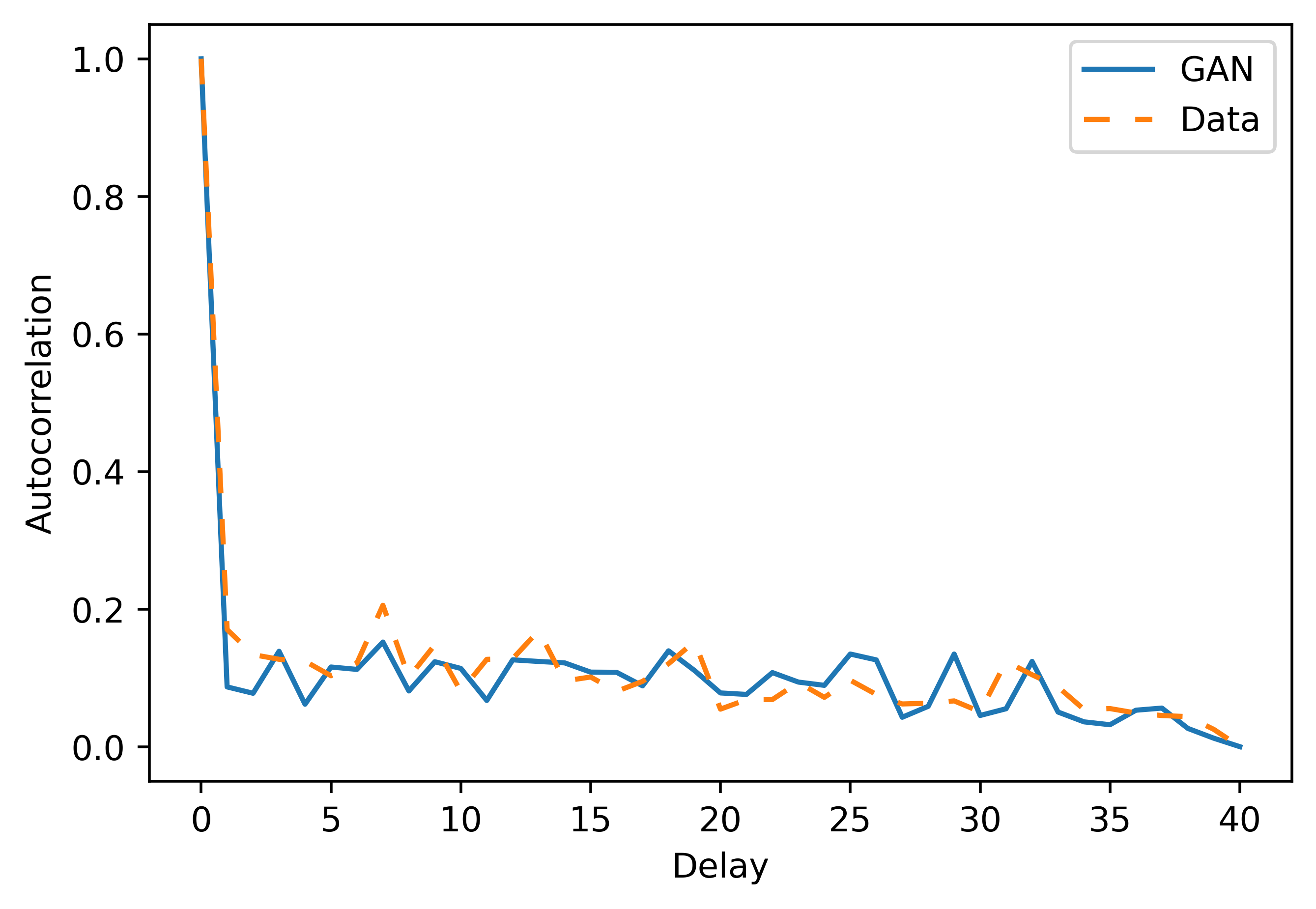}
    \caption{Autocorrelation of a GAN waveform matrix \mbf{X} and training data sample.}
    \label{fig:delay_correlation}
\end{figure}

Figure \ref{fig:gan_bp} presents a variety of beampatterns\footnote{Assuming a uniform linear array with half wavelength inter-sensor spacing.} produced by the GAN model trained on beams 10 to 60 degrees in width in 2 degree increments. The training data also includes a 60 degree beam with a 10 degree space in the middle for avoidance of interfering transmitters. This results in a total of 27 training data classes, not all of which are illustrated here. The training data includes 1000 samples per class. The generator learns to accurately reproduce the beampattern of the different training data classes. The training data 
was generated without constraining the auto and cross-correlations of the waveforms at multiple delays to keep the amount of computation manageable. While they were not optimized during data generation, these properties can be learned implicitly \cite{Paper3,Paper2}. Figure \ref{fig:delay_correlation} shows a representative sample of the autocorrelation magnitude $\abs{\Tr(\mathbf{X}^H\mathbf{X}_{\tau})}$ of GAN and training data waveforms \mbf{X}, where $\mathbf{X}_{\tau}$ is a copy of \mbf{X} delayed by $\tau$ samples.

Waveform diversity is beneficial in many scenarios including dense spectrum use, radar and electronic warfare applications. As we aim to discover new waveforms, it is also essential that the generative model does not directly copy training data. Table \ref{table:Diversity} quantifies diversity in terms of similarity $\abs{\abs{\mathbf{A}^H\mathbf{B}}}_F$ between code matrices \mbf{A} and \mbf{B}. The values are presented for the same model as the beampatterns in Figure \ref{fig:gan_bp}, averaged over 100 generated samples per label. $C_{in}$ represents the similarity $\abs{\abs{\mathbf{X}^H\overline{\mathbf{X}}}}_F$ of a sample $\mathbf{X}$ to the average of all samples $\overline{\mathbf{X}}$ in the set, calculated for both the GAN and the training data. This aims to capture the size of the distribution. $C_{nn}$ represents the (averaged) nearest-neighbor similarity from a GAN synthesized sample to the training data, normalized relative to the respective value of the training data compared to itself. As such, it would be 1 for a model that only directly copies the training dataset.
The GAN does not capture as large a distribution as the training data, but learns to produce novel waveforms nonetheless.

\begin{table}[h!]
\caption{Diversity properties of GAN outputs compared to training data in terms of a similarity between code matrices. $C_{in}$ represents average similarity of a sample in the set to the mean of all samples in the set. $C_{nn}$ represents average nearest-neighbor similarity from a GAN synthesized sample to the training data, normalized such that the value would be 1 for a model that copies the training data directly. GAN produces diverse outputs that are distinct from the training data, but does not achieve the full diversity of the data distribution.}
\centering
\begin{tabular}{||c c||} 
\hline
Model & Value \\ [0.5ex] 
\hline\hline
$C_{in}(\text{GAN})$ & 2.314 \\
$C_{in}(\text{Data})$ & 1.241 \\
$C_{nn}$ & 0.4559 \\ [1ex] 
\hline
\end{tabular}
\label{table:Diversity}
\end{table}

Finally, Table \ref{table:ComputationalTime} presents the time in seconds to generate a single code matrix \mbf{X} for a given \mbf{R}, as well as the generation time for 100 matrices. These values assume that the GAN model has already been trained. The generative model is compared to the optimization-based MultiCAO \cite{Stoica_Synthesis} method to get an idea of the speedup in waveform synthesis offered by the GAN. MultiCAO is run until the change in \mbf{X} between consecutive iterations of the method is below a threshold of 1e-3 (in Frobenius norm). In these results, MultiCAO does not consider delayed copies of the waveform, which is highly favorable for its computational speed. In a more realistic scenario, a greater relative computational speedup would be expected. Furthermore, the execution time of MultiCAO is highly dependent on initialization and \mbf{R}. As such, it is averaged over 100 runs of each of the 27 \mbf{R} used in the generation of our dataset. The generation times for the group of 100 matrices assume the outputs are produced as a single batch using the GAN with the built-in parallelization of PyTorch. 
This generation time for larger groups demonstrates that the computational time for the GAN model does not scale linearly with the number of inputs / outputs in a batch. As such, the computational advantage of the GAN method becomes more significant when producing large numbers of outputs.

\begin{table}[h!]
\caption{Time required to generate code matrix \mbf{X}. The \emph{trained} GAN is greatly faster than conventional optimization methods, even under favorable conditions for the latter.
}
\centering
\begin{tabular}{||c c c||} 
\hline
Model & Single Sample & 100 Samples \\ [0.5ex] 
\hline\hline
MultiCAO  & 0.098  & - \\
GAN (CPU) & 0.008  & 0.074\\
GAN (GPU) & 0.002 & 0.003\\ [1ex] 
\hline
\end{tabular}
\label{table:ComputationalTime}
\end{table}

\section{Conclusion} \label{sec:conclusion}

This paper proposed a generative deep learning model for synthesizing unit-modulus phase-coded waveforms with user-specified beampatterns. The approach uses a GAN model conditioned to produce a group of waveforms with a specific cross-correlation matrix corresponding to a conditioning input. After training, the method produces novel waveforms on-the-fly that accurately approximate the desired beampattern. In future work, it would be interesting to study the generalization of the method to cross-correlation matrices outside the training data. The diversity of outputs could also be improved to better match or exceed the full diversity of the training data.

\printbibliography

\end{document}